\documentclass[aps,prl,twocolumn,superscriptaddress,nofootinbib]{revtex4-2}
\pdfoutput=1
\usepackage[latin9]{inputenc}
\usepackage{graphicx}
\usepackage{amssymb}
\usepackage{amsmath}
\usepackage{color}
\usepackage{bm}
\usepackage{dsfont}
\usepackage{xspace}
\usepackage{capt-of}
\usepackage[colorlinks=true,urlcolor=blue,linkcolor=blue,citecolor=blue,bookmarks=false]{hyperref}

\newcounter{myequation}
\makeatletter
\@addtoreset{equation}{myequation}
\makeatother

\newcounter{myfigure}
\makeatletter
\@addtoreset{figure}{myfigure}
\makeatother

\newcommand{\qph}{\quad \phantom{.}}

\newcommand{\w}{\omega}

\newcommand{\ad}{\alpha_\mathrm{d}^\ast}
\newcommand{\mloc}{m_\mathrm{loc}}

\newcommand{\im}{\mathrm{i}}

\newcommand{\customref}[2]{\hyperref[#1]{\ref*{#1}#2}}

\definecolor{darkgreen}{rgb}{0,0.5,0}
\definecolor{darkblue}{rgb}{0,0,0.5}
\definecolor{purple}{rgb}{0.35,0,0.35}
\definecolor{orange}{rgb}{1,0.5,0}

\begin{document}
\title{
SU(2)-Symmetric Spin-Boson Model:\\
Quantum Criticality, Fixed-Point Annihilation, and Duality
}

\author{Manuel Weber}
\affiliation{Max-Planck-Institut f\"ur Physik komplexer Systeme, N\"othnitzer Str.~38, 01187 Dresden, Germany}
\affiliation{Institut f\"ur Theoretische Physik and W\"urzburg-Dresden Cluster of Excellence ct.qmat, Technische Universit\"at Dresden, 01062 Dresden, Germany}
\author{Matthias Vojta}
\affiliation{Institut f\"ur Theoretische Physik and W\"urzburg-Dresden Cluster of Excellence ct.qmat, Technische Universit\"at Dresden, 01062 Dresden, Germany}

\date{\today}

\begin{abstract}
The annihilation of two intermediate-coupling renormalization-group (RG) fixed points is of interest in diverse fields from statistical mechanics to high-energy physics,
but has so far only been studied using perturbative techniques.
Here we present high-accuracy quantum Monte Carlo results for the SU(2)-symmetric $S=1/2$ spin-boson (or Bose-Kondo) model. We study the model with a power-law bath spectrum $\propto \omega^s$ where, in addition to a critical phase predicted by perturbative RG, a stable strong-coupling phase is present. Using a detailed scaling analysis, we provide direct numerical evidence for the collision and annihilation of two RG fixed points at $s^\ast = 0.6540(2)$, causing the critical phase to disappear for $s<s^\ast$.
In particular, we uncover a surprising duality between the two fixed points, corresponding to a reflection symmetry of the RG beta function, which we utilize to make analytical predictions at strong coupling which are in excellent agreement with numerics.
Our work makes phenomena of fixed-point annihilation accessible to large-scale simulations, and we
comment on the consequences for impurity moments in critical magnets.
\end{abstract}

\maketitle


A comprehensive understanding of critical phenomena requires one to unravel their underlying renormalization-group (RG) structure. A scenario of current interest is the collision and annihilation of two intermediate-coupling fixed points as a function of an external parameter, such as spatial dimension, as it leads to the breakdown of naive perturbative predictions and to an unconventionally slow RG flow close to the collision, associated with nontrivial crossover phenomena \cite{kaplan09}. Fixed-point collisions have been discussed in a number of contexts using RG techniques, including the Abelian Higgs model \cite{halperin74,braun14}, the chiral phase transition in quantum chromodynamics \cite{gies06,kaplan09}, the $Q$-state Potts model \cite{Nienhuis79}, and deconfined criticality in quantum magnets \cite{nahum15,wang17,Nahum20}. Detailed numerical studies are lacking, however, partly because dimensionality cannot be tuned continuously.

Dissipative quantum impurity models play a central role in modern physics, with applications from fundamental statistical mechanics to biological systems \cite{leggett}.
In condensed matter, they can serve as effective models for magnetic moments in magnets \cite{SBV99,VBS00}, but also for heavy-fermion metals described by extended dynamical mean-field theory (EDMFT) \cite{si01,si03} as well as non-Fermi-liquid behavior
in Sachdev-Ye-Kitaev-type (SYK-type) models \cite{SY93,SYKreview}.
Already simple dissipative impurity models can show nontrivial quantum phase transitions \cite{leggett, mv06}, and they have recently been proposed as platform to access fixed-point annihilation numerically \cite{guo12,bruo14}, but direct evidence is missing.

It is the purpose of this Letter to close this gap. We utilize a recently developed wormhole quantum Monte Carlo (QMC) technique that enables us to simulate dissipative quantum systems to far lower temperatures than previously possible \cite{weber21}. We study an SU(2)-symmetric three-bath generalization of the $S=1/2$ spin-boson model \cite{bfk,SBV99,VBS00,sengupta}, with a gapless bath spectrum scaling as $\w^s$. For this model, we firmly establish the existence of both critical and strong-coupling phases and obtain precise values for the critical exponents at the associated transition.
The critical phase disappears from the phase diagram for $s<s^\ast = 0.6540(2)$ due to a collision of RG fixed points. Using a scaling analysis, we directly monitor this fixed-point collision in an unprecedented manner. Most importantly, we find a remarkable duality between the stable and unstable fixed points located at small and large intermediate coupling, respectively. This enables us to draw conclusions on the nature of the strong-coupling expansion and to deduce exact results near $s^\ast$ even in the absence of a small parameter. Our work paves the way to high-accuracy studies of more complex dissipative quantum models and, for the first time, makes the fixed-point annihilation accessible to large-scale simulations.

\textit{Model and phase diagram.}---%
The standard spin-boson model, where a spin-$1/2$ is subjected to a transverse field and coupled to a single bosonic bath, features a quantum phase transition between a delocalized and a localized phase \cite{kehrein96,bulla03}. This transition obeys a quantum-to-classical correspondence (QCC) \cite{VTB05,rieger09,guo12}, i.e., is in the same universality class as the thermal phase transition of the one-dimensional Ising model with $1/r^{1+s}$ interactions \cite{fisher_critical_1972}.
Generalizations of the model to multiple baths have been found to show more complex behavior. In particular, the existence of a nonclassical critical phase has been predicted using perturbative RG and large-$N$ techniques \cite{bfk,SBV99,VBS00,sengupta}. For the two-bath case, extensive numerical results have been obtained using matrix-product-state (MPS) techniques \cite{guo12,bruo14}; they confirmed the absence of QCC, but also signaled that the perturbative prediction becomes invalid for small $s$ where only a strong-coupling phase was shown to exist. For the more relevant SU(2)-symmetric three-bath case, high-accuracy results are lacking, as established numerical methods such as MPS \cite{guo12} and Wilson's numerical renormalization group \cite{bulla03,bulla05} become prohibitively expensive. Consequently, relatively little precise information is available on the phases and transitions of this model \cite{otsuki13,cai19}. This is pressing since a number of conclusions drawn in earlier work rely on extrapolating the perturbative $\epsilon$-expansion results to $\epsilon = 1-s\to 1$ \cite{SBV99,VBS00,si01,si03}.

\begin{figure}
\includegraphics[width=\columnwidth]{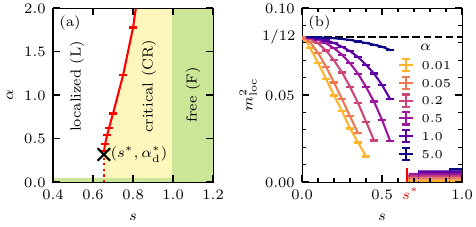}
\caption{%
(a)~Phase diagram as a function of the bath exponent $s$ and the spin-boson coupling $\alpha$.
The line of second-order quantum phase transitions (red) between the CR and L phases terminates at the point $(s^\ast,\ad)$, such that the CR phase disappears for $s<s^\ast$; for details see text.
(b)~Local moment $\mloc^2 = \lim_{\beta\to\infty} \langle S^x(\beta/2) S^x(0) \rangle$ for different $\alpha$. We only show data that are converged in temperature.
For each $\alpha$, the corresponding horizontal line in the lower right corner marks the onset of the CR phase where $\mloc^2 = 0$ for $s>s(\alpha_\mathrm{QC})$.
The dashed line corresponds to $\mloc^2(\alpha\to\infty)=S^2/3=1/12$.
}
\label{fig:phd}
\end{figure}

In this Letter, we consider this three-bath generalization of the spin-boson model,
\begin{eqnarray}
\label{eq:h}
\mathcal{H}
&=&
   \sum_{i = x,y,z} \sum_{q} \left[
\lambda_{q i} S^i
( \hat{B}_{q i} + \hat{B}_{q i}^{\dagger} )
+
\w_q \hat{B}_{q i}^{\dagger} \hat{B}_{q i}
\right] .
\qph
\end{eqnarray}
Here, an $S=1/2$ spin $\vec{S}$ is coupled to three independent bosonic baths, whose spectral densities $J_i(\w) = \pi \sum_{q} \lambda_{q i}^{2} \delta(\w -\w_q)$ are
of power-law form,
\begin{equation}
J_i(\w) = 2\pi\, \alpha_i\, \omega_{\rm c}^{1-s} \, \omega^s\,, \quad
 0<\w<\w_{\rm c} =1 \, ,
\label{power}
\end{equation}
with $\omega_{\rm c}$ being a cutoff energy. The model \eqref{eq:h} displays nontrivial quantum dynamics even in the absence of an additional external field: The noncommuting character of the three spin components implies that the localization tendencies induced by each bath compete. The $\alpha_i$ measure the dissipation strength, and we focus on the SU(2)-symmetric case $\alpha_i\equiv\alpha$.

Figure~\ref{fig:phd}(a) displays the model's quantitative phase diagram obtained from our QMC simulations. For bath exponents $s^\ast<s<1$, both critical (CR) and localized (L) phases exist and are separated by a continuous quantum phase transition. In contrast, for $s>1$ the model is always in a free-spin (F) phase, whereas for $s<s^\ast$ any finite coupling drives the system into the L phase. This L phase exhibits spontaneous breaking of the SU(2) symmetry, whereas the F phase features an asymptotically free spin. Finally, the CR phase is characterized by fractional power-law spin correlations \cite{SY93,SBV99,VBS00}, $\chi_i(\w) \propto \w^{-x}$ where $\chi_i(\tau) = \langle S^i(\tau) S^i(0)\rangle \sim 1/\tau^{1-x}$.


\textit{Perturbative RG and fixed-point collision.}---%
Before we discuss the details of our numerical results, we review what is known about the fixed-point structure from
weak-coupling perturbative RG \cite{VBS00,rg_bfk}. The impurity--bath coupling is marginal at $s=1$, and expanding about the $\alpha=0$ free-spin fixed point results in the two-loop beta function \cite{rg_bfk,ren_note},
\begin{align}
\label{beta_wk}
\beta(\alpha) \equiv \frac{d\alpha}{d\ln\mu} &= -(1-s) \alpha + 4 \alpha^2 - 8 \alpha^3 \, ,
\end{align}
where $\mu$ is the RG reference scale. From this beta function, one deduces the existence of an infrared-stable fixed point for $s<1$, located at
\begin{equation}
\label{alstar}
\alpha_{\rm CR}^\ast = \frac{1-s}{4} + \frac{(1-s)^2}{8} + \mathcal{O}[(1-s)^3] \,;
\end{equation}
this is the CR fixed point corresponding to the CR phase.
Its properties can be obtained in a double expansion in $\alpha$ and $(1-s)$. The power-law spin autocorrelations are characterized by the exponent $x$, and $x=s$ is an exact result following from the diagrammatic structure of the susceptibility \cite{VBS00,rg_bfk}.

\begin{figure}
\includegraphics[width=\columnwidth]{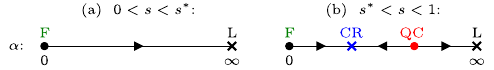}
\caption{%
Schematic RG flow for $0<s<s^\ast$ and $s^\ast < s <1$.
}
\label{fig:RG}
\end{figure}

The two-loop beta function in Eq.~\eqref{beta_wk} does display {\em two} nontrivial fixed points at
$\alpha^\ast_{1,2} = \frac{1}{4} [1 \pm \sqrt{1 - 2(1-s)}]$,
with $\alpha^\ast_2$ being the stable CR fixed point of Eq.~\eqref{alstar}, and $\alpha^\ast_1$ being infrared unstable. These two fixed points approach each other upon decreasing $s$ from unity, such that they collide as $s\to s^{\ast,+}$, with $s^\ast=1/2$ from \eqref{beta_wk}. Although $\alpha^\ast_1$ is outside the regime of validity of the epsilon expansion, the comparison with numerics suggests to identify $\alpha^\ast_1$ with the quantum critical (QC) fixed point controlling the transition between the CR and L phases.

The resulting schematic RG flow is depicted in Fig.~\ref{fig:RG}: Two intermediate-coupling fixed points exist for $s^\ast<s<1$, while they disappear for $s<s^\ast$ leaving only runaway flow to strong coupling. As we show below, our numerical results not only prove this picture to be correct but also indicate a surprising duality between the CR and QC fixed points;
below, we will use this duality with Eqs.~\eqref{beta_wk} and \eqref{alstar} to make analytical predictions at strong coupling.


\textit{QMC method.}---%
For our simulations, we used an exact QMC method with global wormhole updates \cite{weber21}
that samples a diagrammatic expansion of the partition function in the retarded spin interaction \cite{weber17},
originating from tracing out the bosonic bath in Eq.~\eqref{eq:h} analytically.
A detailed description of our method can be found in Ref.~\cite{weber21}.


\begin{figure}
\includegraphics[width=\columnwidth]{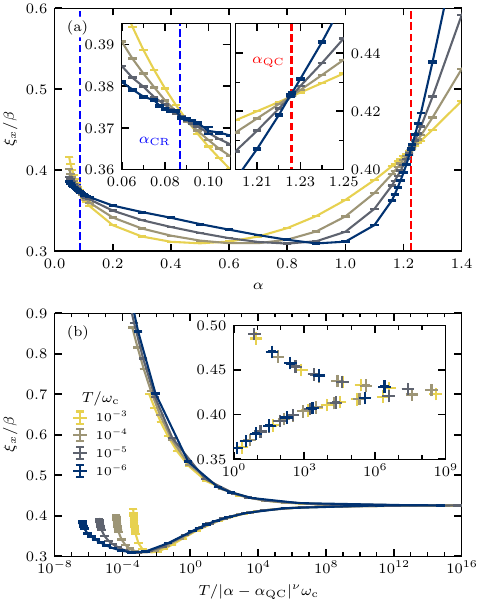}
\caption{%
(a)~Correlation length $\xi_x/\beta$ as a function of $\alpha$ for different temperatures and $s=0.75$.
The insets show close-ups of the crossings near the CR and QC fixed points.
(b)~Data collapse for $\xi_x/\beta$ near the QC fixed point. The inset shows a detailed view of the
critical region. Finite-size scaling yields $1/\nu=0.192(2)$
\cite{SupplInfo}.
}
\label{fig:scal1}
\end{figure}

\textit{QMC results.}---%
To determine the phase diagram in Fig.~\ref{fig:phd}(a), we calculate the dynamical spin susceptibility
$
\chi_x(\im\Omega_n)
	=
	\int_0^{\beta} d\tau \, e^{\im \Omega_n \tau} \left\langle S^x(\tau) S^x(0) \right\rangle
$
from the imaginary-time spin correlations in the $x$ direction. Here, $\Omega_n = 2\pi n / \beta$, $n\in\mathds{Z}$, are the bosonic Matsubara frequencies and $\beta=1/T$ is the inverse temperature.
The different phases in Fig.~\ref{fig:phd}(a) can be identified from the low-temperature behavior of the static susceptibility $\chi_x \equiv \chi_x (\im \Omega_0)$:
At $\alpha=0$, $\chi_x(T) = 1/ (4T)$.
In the F and L phases, $\chi_x(T) = \mloc^2 / T$ follows a Curie law as $T\to0$ with
a reduced but finite local moment;
the latter is determined as
$\mloc^2 = \lim_{\beta\to\infty} \langle S^x(\beta/2) S^x(0) \rangle$
and shown in Fig.~\ref{fig:phd}(b).
We find that $\mloc^2(s \to 0)$ approaches
the strong-coupling result $\mloc^2(\alpha\to\infty) = S^2 /3 = 1/12$ of the L phase
independent of $\alpha$; this is the local moment of a classical spin.
In the CR phase, $\mloc^2=0$ and $\chi_x(T) \propto T^{-s}$.
For further details, see the Supplemental Material \cite{SupplInfo}.

For a quantitative analysis of criticality, we consider the correlation length along the imaginary-time axis
(correlation time),
$
\xi_x
	=
	\frac{1}{\Omega_1} \sqrt{\frac{\chi_x(\im\Omega_0)}{\chi_x(\im\Omega_1)}-1}
$.
In analogy to the definition of the spatial correlation length \cite{SupplInfo, Sandvik10}, we identify the inverse temperature  $\beta$ with the system size in the imaginary-time direction and the Matsubara frequency $\Omega_0$ with the ordering vector [note that the lowest Matsubara frequencies determine the long-time decay of $\chi_x(\tau)$]. Figure \ref{fig:scal1}(a) depicts $\xi_x / \beta$ as a function of $\alpha$ for $s=0.75$ and different temperatures. We find that $\xi_x / \beta$ diverges in the L phase, but remains finite in the CR phase.
Because $\xi_x / \beta$ becomes an RG-invariant quantity at the fixed points of the beta function, we can identify the two sharp crossings observed in the insets of Fig.~\ref{fig:scal1}(a) with the corresponding fixed-point couplings $\alpha_\mathrm{CR}$ and $\alpha_\mathrm{QC}$.
Moreover, Fig.~\ref{fig:scal1}(b) demonstrates that $\xi_x / \beta$ fulfills the scaling ansatz
\begin{align}
\label{eq:corrlen}
\xi_x / \beta =
	f \boldsymbol{(}\beta^{1/\nu} (\alpha - \alpha_\mathrm{QC})\boldsymbol{)}
\end{align}
near the
critical coupling $\alpha_\mathrm{QC}$. Here, $\nu$ is the correlation-length exponent and $f$ is
a universal function. We observe a clear data collapse over many orders of magnitude in the energy scale $T/|\alpha - \alpha_\mathrm{QC}|^\nu$ and for temperatures $T/\omega_\mathrm{c} \lesssim 10^{-3}$. Details on how we estimate $\alpha_\mathrm{CR}$, $\alpha_\mathrm{QC}$, and $\nu$ are provided in the Supplemental Material \cite{SupplInfo}.

\begin{figure}
\includegraphics[width=\columnwidth]{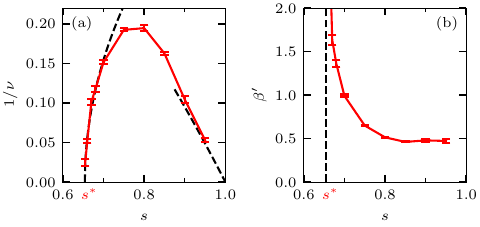}
\caption{%
(a)~Inverse correlation-length exponent $1/\nu$ associated with the QC fixed point as a function of $s$.
$\nu$ diverges for both $s\to 1$ and $s\to s^\ast$. The dashed lines show the predictions \eqref{eq:nu1} and \eqref{eq:nu2} based on fixed-point duality; for the latter, we fit
$\sqrt{A_1 B_0}=0.72(2)$.
(b) Magnetization exponent $\beta'$ calculated from $1/\nu$ via the hyperscaling relation $\beta'/\nu = (1-s)/2$.
}
\label{fig:exp}
\end{figure}

Figure~\ref{fig:exp} shows the critical exponents as a function of $s$. The correlation-length exponent $\nu$ estimated from Eq.~(\ref{eq:corrlen}) diverges for both $s \to 1$ and $s\to s^\ast$ [see Fig.~\ref{fig:exp}(a)] and the leading behavior is consistent with the predictions \eqref{eq:nu1} and \eqref{eq:nu2} derived below. As demonstrated in the Supplemental Material \cite{SupplInfo}, the remaining exponents are completely determined by hyperscaling relations.
Close to criticality, the local moment
fulfills $\mloc \propto (\alpha - \alpha_\mathrm{QC})^{\beta'}$.
The magnetization exponent $\beta'$ is summarized in Fig.~\ref{fig:exp}(b); it approaches $\beta'\approx 1/2$ for $s\to 1$ but diverges for $s \to s^\ast$.
This divergence results from the fixed-point collision and leads to an extremely slow RG flow for intermediate $\alpha$ and $s\sim s^\ast$ \cite{kaplan09}. In particular, the order parameter $\mloc$ is exponentially suppressed \cite{SupplInfo} in the region $\alpha < \ad$ and $s\lesssim s^\ast$ in Fig.~\ref{fig:phd}(b), such that a naive extrapolation of $\mloc^2$ to zero would significantly underestimate $s^\ast$.
This is likely the reason why the value of $s^\ast$ as estimated in Ref.~\cite{cai19} significantly deviates from ours.
We also note that such a fixed-point collision is not present in any of the relevant classical spin models, hence QCC is violated for the model under consideration.

\begin{figure}
\includegraphics[width=\columnwidth]{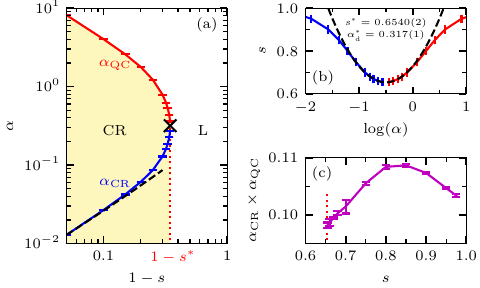}
\caption{%
Fixed-point duality.
(a)~Location of the two intermediate-coupling fixed points CR and QC, as determined from crossing points of $T^s \chi_x$ \cite{SupplInfo}, as a function of the bath exponent $s$. The black dashed line indicates the prediction (\ref{alstar}) of the perturbative RG for $\alpha_\mathrm{CR}$.
(b)~Close to $s^\ast$, the fixed-point collision is well approximated by $s = s^\ast + \frac{B_0}{A_1} \ln^2(\alpha/\ad)$ from which
we extract $s^\ast = 0.6540(2)$, $\ad = 0.317(1)$ where the fixed points disappear, and $A_1/B_0 = 17.7(2)$.
(c)~Product of the two fixed-point couplings as a function of $s$: This is approximately constant despite
the couplings
varying over several orders of magnitude.
}
\label{fig:duality}
\end{figure}

Figure \ref{fig:duality}(a) shows the evolution of $\alpha_\mathrm{CR}$ and $\alpha_\mathrm{QC}$ as a function of $(1-s)$. For small $(1-s)$, the former closely follows the RG prediction (\ref{alstar}), whereas the latter diverges proportional to $1/(1-s)$. Remarkably, the evolution of  $\alpha_\mathrm{CR}$ and $\alpha_\mathrm{QC}$ as a function of $(1-s)$ is almost symmetric in $\ln(\alpha)$ until they coalesce at $(s^\ast, \ad)$.
The fixed-point collision appears at $s^\ast = 0.6540(2)$ and $\ad = 0.317(1)$ which we estimate from
a quadratic fit in $\ln(\alpha)$, as shown in Fig.~\ref{fig:duality}(b).
As a function of $s$, the product $\alpha_\mathrm{CR} \times \alpha_\mathrm{QC}$ in Fig.~\ref{fig:duality}(c) varies by around 10$\%$ which is approximately constant considering that each coupling varies over several orders of magnitude.


\textit{Duality.}---%
The data in Fig.~\ref{fig:duality} show that the two zeros of the (exact) beta function are located symmetrically with respect to $\ln(\ad)$ for all $s$ to very good accuracy, hence $\alpha_{\rm QC} /\ad = \ad / \alpha_{\rm CR}$. We conjecture that this symmetry is obeyed by the full beta function, $(1/\alpha)\beta(\alpha) = \tilde\beta(\alpha/\ad)$ with $\tilde\beta(x) = \tilde\beta(1/x)$, see Fig.~\ref{fig:beta1}. This implies a duality between the two fixed-point theories which we discuss in the following.

First, the properties of QC near $s=1$ can be deduced from the dual of the weak-coupling expansion \eqref{beta_wk}; we note that a suitable field theory for that is not known. Introducing $\bar\alpha\equiv {\ad}^2/\alpha$, we have by duality
\begin{align}
\label{beta_str}
\beta(\bar\alpha) &= (1-s) \bar\alpha - 4\bar\alpha^2 + 8\bar\alpha^3\,,
\end{align}
where the sign change compared to \eqref{beta_wk} arises from $d\ln\alpha/d\ln\mu = -d\ln(1/\alpha)/d\ln\mu$. The QC fixed point is then located at
\begin{equation}
\label{eq:alqc}
\bar\alpha_{\rm QC}^\ast = \frac{1-s}{4} + \frac{(1-s)^2}{8} + \mathcal{O}[(1-s)^3]\,,
\end{equation}
and its correlation-length exponent is obtained from expanding $\beta(\bar\alpha)$ about $\bar\alpha_{\rm QC}^\ast$, resulting in
\begin{equation}
\label{eq:nu1}
1/\nu = 1-s - \frac{(1-s)^2}{2} + \mathcal{O}[(1-s)^3] \, .
\end{equation}

\begin{figure}
\includegraphics[width=\columnwidth]{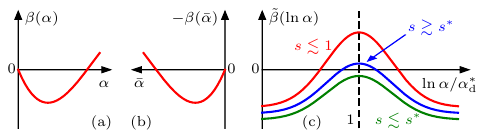}
\caption{%
(a)~Weak-coupling and (b) its dual strong-coupling beta function,
as in Eqs.~(\ref{beta_wk}) and (\ref{beta_str}).
(c)~Reflection symmetry of $\tilde{\beta}(\ln \alpha)$ around $\alpha^\ast_\mathrm{d}$.
The zeros of $\beta(\alpha)$ disappear for $s<s^\ast$.
}
\label{fig:beta1}
\end{figure}

Second, to study the fixed-point annihilation at $s^\ast$, we expand the beta function near $\ad$ as $\beta(\alpha) = A(s) - B(s) \ln^2(\alpha/\ad)$ \cite{kaplan09}, with $A(s) = A_1 (s-s^\ast)$ and $B(s) = B_0(s^\ast) + B_1 (s-s^\ast)$.
This yields the locations of the two fixed points as $\pm\ln(\alpha^\ast/\ad)=\sqrt{A_1/B_0}\sqrt{s-s^\ast}$ and the correlation-length exponent of QC as
\begin{equation}
\label{eq:nu2}
1/\nu = \sqrt{A_1 B_0} \sqrt{s-s^\ast} \, .
\end{equation}
Both predictions \eqref{eq:nu1} and \eqref{eq:nu2} are in good agreement with the QMC data in Fig.~\ref{fig:exp}.

While Eq.~\eqref{eq:nu2} generically applies near a fixed-point collision, Eqs.~\eqref{eq:alqc} and \eqref{eq:nu1} rely on the conjectured mirror symmetry of the beta function. Our numerics indicate that this symmetry is not exact, hence the weak- and strong-coupling expansions possibly differ in higher loop orders.


\textit{Impurities in quantum critical magnets.}---%
Previous work \cite{SBV99,VBS00} on magnetic impurities in quantum critical magnets in $d=3-\epsilon$ space dimensions---a problem closely related to the three-bath spin-boson model---employed a weak-coupling expansion similar to that in Eq.~\eqref{beta_wk}, with the difference that interactions among the bath bosons are RG relevant for $d<3$. The physically most interesting case of $d=2$ corresponds to a bath exponent of $s=0$. References \cite{SBV99,VBS00} assumed continuity from small $\epsilon=1-s$ to $\epsilon=1$, and numerical results have been interpreted in terms of the CR fixed point \cite{sandvik07}. Given that this continuity does {\em not} hold for the noninteracting bath case studied here, the present results raise the interesting question whether a strong-coupling phase also occurs for the impurity-in-a-magnet problem and, if yes, what its properties are.

Relatedly, conclusions which were drawn from $\epsilon$ or large-$N$ expansion results for quantum critical lattice models in the framework of
EDMFT \cite{si01,si03,cai19} need to be revisited.


\textit{Conclusions.}---%
Using high-accuracy QMC simulations, we have determined the phase diagram and critical properties of the SU(2)-symmetric spin-boson model. For the first time, we were able to extract the location of both intermediate-coupling fixed points using a scaling analysis and to monitor their collision and subsequent annihilation as a function of the bath exponent $s$. The fixed points display a remarkable duality, which we have utilized to deduce analytical results at strong coupling; we hope that future studies will give further insight into this novel duality relation.
Our results illustrate the power of the QMC algorithm of Ref.~\cite{weber21},
which makes the analysis of unconventionally slow RG flow near the collision \cite{kaplan09} accessible to future numerical studies.
In the context of SYK models, it has been suggested that the localized phase of the single-impurity problem triggers the existence of an SYK spin-glass state \cite{SYKreview}; the role of the fixed-point annihilation is an interesting open problem.

\begin{acknowledgments}
We thank F.~Parisen Toldin for helpful discussions.
This work has been supported by the Deutsche Forschungsgemeinschaft through the W\"urzburg-Dresden Cluster of Excellence on Complexity and Topology in Quantum Matter -- \textit{ct.qmat} (EXC 2147, Project No.\ 390858490) and SFB 1143 on Correlated Magnetism (Project No.\ 247310070).
\end{acknowledgments}

\textit{Note added.}---%
Recently, we became aware of Refs.~\cite{cuoma22,beccaria22,Nahum22} in which the fixed-point annihilation
is studied analytically in the large-$S$ limit.
These results indicate that the duality relation of the beta function becomes exact for
$S\to\infty$, but this is not discussed in Refs.~\cite{cuoma22,beccaria22,Nahum22}.


\clearpage

\stepcounter{myequation}
\stepcounter{myfigure}

\setcounter{secnumdepth}{3}  

\renewcommand{\thefigure}{S\arabic{figure}}
\renewcommand{\thesection}{S\arabic{section}}
\renewcommand{\thetable}{S\arabic{table}}
\renewcommand{\theequation}{S\arabic{equation}}

\newcommand{\bmin}{\beta_\mathrm{min}}
\newcommand{\bmax}{\beta_\mathrm{max}}
\newcommand{\nmax}{n_\mathrm{max}}
\newcommand{\alphac}{\alpha_\mathrm{c}}

\newcommand{\ie}[0]{i.e.\@\xspace}
\newcommand{\eg}[0]{e.g.\@\xspace}
\newcommand{\etal}[0]{\textit{et al.}\@\xspace}
\newcommand{\cf}[0]{cf.\@\xspace}
\newcommand{\ia}[0]{i.a.\@\xspace}

\definecolor{Gray}{gray}{0.9}

\onecolumngrid

\centerline{\bf\large Supplemental Material for} \vskip3mm
\centerline{\bf\large SU(2)-Symmetric Spin-Boson Model:} \vskip0.5mm
\centerline{\bf\large Quantum Criticality, Fixed-Point Annihilation, and Duality} \vskip1cm

\twocolumngrid

\author{Manuel Weber}
\affiliation{Max-Planck-Institut f\"ur Physik komplexer Systeme, N\"othnitzer Str.~38, 01187 Dresden, Germany}
\affiliation{Institut f\"ur Theoretische Physik and W\"urzburg-Dresden Cluster of Excellence ct.qmat, Technische Universit\"at Dresden,
01062 Dresden, Germany}
\author{Matthias Vojta}
\affiliation{Institut f\"ur Theoretische Physik and W\"urzburg-Dresden Cluster of Excellence ct.qmat, Technische Universit\"at Dresden,
01062 Dresden, Germany}

\date{\today}

\maketitle

Here we present supplementary information
for the results discussed in our main paper. In Sec.~\ref{Sec:LocMom}, we
expand on our discussion of the local moment: we show how
it is extracted from our finite-temperature data and present
additional results in the F phase. In Sec.~\ref{Sec:FSS}, we
explain the details of our finite-size-scaling analysis to extract
the fixed-point couplings $\alpha_\mathrm{CR}$ and
$\alpha_\mathrm{QC}$ as well as the inverse correlation-length
exponent $1/\nu$. In Sec.~\ref{Sec:hyperhyper}, we show
that the remaining critical exponents are fully determined by hyperscaling
relations.

\section{Formation of the local moment\label{Sec:LocMom}}

The local moment determines the asymptotic behavior of the spin susceptibility
\begin{align}
\chi_x = 
	\frac{m^2_\mathrm{loc}}{T}
	\quad \mathrm{for} \quad T\to 0
	\, .
\end{align}
From our finite-temperature data, it is more convenient to extract $m_\mathrm{loc}^2$
from the imaginary-time spin-spin correlation function at distance $\tau = \beta/2$,
\begin{align}
\label{Eq:orderp}
m^2_\mathrm{loc}(T)
	=
	\langle S^x(\beta/2) S^x(0) \rangle \, ,
\end{align}
and then take the limit $m^2_\mathrm{loc} = m^2_\mathrm{loc}(T\to0)$, because
$m^2_\mathrm{loc}(T)$ converges faster than $T\chi_x$.

\begin{figure}[ht!]
\includegraphics[width=0.97\columnwidth]{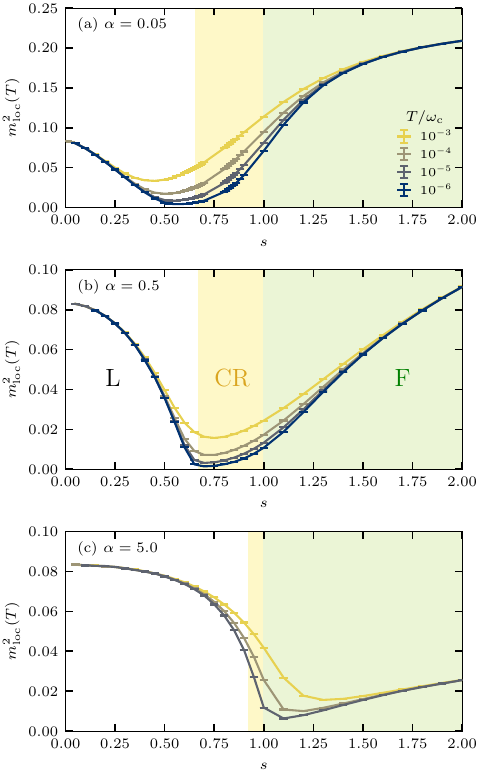}
\caption{%
Convergence of the finite-temperature estimator of the local moment, $m^2_\mathrm{loc}(T)$,
as a function of $s$ and for different $\alpha$. The local moment is finite in the L and F phases,
but scales to zero in the CR phase.
}
\label{fig:localmom}
\end{figure}

Figure \ref{fig:localmom} shows the finite-temperature convergence of $m^2_\mathrm{loc}(T)$
for different couplings $\alpha$ as a function of the bath exponent $s$. We observe that
deep in the L and F phases, $m^2_\mathrm{loc}(T)$ has already converged to its ground-state
value for the temperatures available in our simulations.
When approaching $s \to s^{\ast,-}$ from the L phase, $m^2_\mathrm{loc}$ becomes exponentially small
for $\alpha < \alpha_\mathrm{d}^\ast$, as indicated for $\alpha=0.05$ in Fig.~\ref{fig:localmom}(a).
As discussed in our main paper, this is a consequence of the diverging $\beta'$ exponent
and a characteristic property of the slow RG flow close to the fixed-point annihilation \cite{kaplan09}.
In the CR phase, $m^2_\mathrm{loc}(T)$
still shows substantial temperature effects and only slowly vanishes. For small $\alpha$,
convergence in the CR phase is slowest for $s\to 1^-$ because the exponent
of the expected power-law behavior in $\chi_x \propto T^{-s}$ becomes hard to distinguish
from a Curie law. If we approach the limit $s\to1^+$ from the F phase, $m^2_\mathrm{loc}(s)$
appears to continuously approach zero. This suggests that at $s=1$ the susceptibility has multiplicative corrections to the expected Curie-law behavior, in accordance with the fact that the coupling to the ohmic bath is a marginally irrelevant perturbation.

The local moment shows very distinct behavior in the strong-coupling limit of the L and F phases.
In the L phase, $m^2_\mathrm{loc}(0<s<1, \, \alpha \to \infty) = S^2 / 3 = 1/12$; for $s\to0$ this result is true for any finite $\alpha$. $m^2_\mathrm{loc}=1/12$ corresponds to the Curie response of a classical spin $1/2$; this can be be rationalized by noting that the local spin becomes locked to the bath creating a large classical object, which rigidly rotates in response to the applied field \cite{SBV99}.
By contrast, Fig.~\ref{fig:localmom} suggests
$m^2_\mathrm{loc}(s>1, \, \alpha \to \infty) = 0$ in the F phase.
Furthermore, for any $0<\alpha<\infty$ we expect
$m^2_\mathrm{loc}(s\to\infty) = S(S+1)/3 = 1/4$ because the spectrum of the bosonic bath gets
depleted at low frequencies, so that we are left with a free spin again.

\section{Finite-size scaling\label{Sec:FSS}}

Finite-size scaling (FSS) is a powerful technique to analyze the critical behavior around
a critical point when data is only available for finite system sizes. As the correlation
length $\xi$ becomes of the order of the system size $L$ at the critical point,
the FSS limit extrapolates towards the thermodynamic limit by taking $\xi, L \to \infty$
at fixed $\xi / L$. Originally, FSS had been formulated for thermal phase transitions of classical systems,
where $L$
describes the spatial extent of the system and $\xi$ the spatial correlation length.
At quantum phase transitions, quantum fluctuations drive a zero-temperature phase transition.
From quantum statistical mechanics, it is well known that the path-integral formulation of the partition
function of a quantum system in $d$ spatial dimensions involves an action in $(d+1)$ space-time dimensions;
the additional dimension corresponds to imaginary time, and the length in this direction is given by the
inverse temperature $\beta=1/T$. Therefore, one can define a correlation length in imaginary time the same way as in real space, resulting in a spatial and a temporal correlation length (the latter also called correlation time). In general, both diverge upon approaching a continuous phase transition, and they are related by the dynamical critical exponent $z$. In order to perform FSS at a quantum phase transition,
one has to ensure that the ratios of both spatial and temporal correlation lengths with respect to the relevant length scales of the
system remain fixed. For the quantum impurity system considered in this work, the spatial dimension is zero, so that we can only extract
the critical behavior from the correlation length in imaginary time. In the same way as for spatially extended systems, we analyze
correlation functions to extract the critical behavior, but now in imaginary time instead of space; in practice, we can easily
transfer the estimators used for FSS from real space to imaginary time, we just have to identify the system size $L$ with the
inverse temperature $\beta$.
A detailed description of FSS at a quantum phase transition
can be found in Ref.~\onlinecite{FSS14}; the notation used below follows Ref.~\onlinecite{toldin15}.

In the following, we use FSS to determine critical couplings and critical exponents.
For our analysis, we consider RG-invariant observables because
they have an advantageous scaling behavior. Near criticality,
they obey the scaling ansatz
\begin{align}
\label{Eq:Rsc}
R(\alpha,\beta)
	&=
	f_R(w) + \beta^{-\omega} g_R(w) \, ,
\\
w &= \left(\alpha - \alphac \right) \beta^{1/\nu} \, ,
\end{align}
where $f_R(w)$ is a universal function in $w$ which only depends on the distance from
the critical coupling $\alphac$ and the correlation-length exponent $\nu$.
We also included the generic form of a subleading
correction, $\beta^{-\omega} g_R(w)$, with exponent $\omega$.
From Eq.~(\ref{Eq:Rsc}) we can determine both $\alphac$ and $\nu$.
If we only want to extract the critical coupling, the crossing method
is often more advantageous: In the absence of scaling corrections,
Eq.~(\ref{Eq:Rsc}) takes a universal value $R^\ast(\alphac)$ independent of $\beta$.
For a pair of temperatures $(\beta, c\, \beta)$ with $c>1$, we have
$R(\alphac,\beta) = R(\alphac,c\,\beta)$, \ie, the two data sets show a crossing
at $\alphac$. Finite scaling corrections will lead to a drift of the crossing, therefore
we define the pseudocritical coupling $\alpha_{\mathrm{c},R}(\beta)$
from
\begin{align}
\label{Eq:Rcross}
R(\alpha_{\mathrm{c},R}(\beta),\beta) = R(\alpha_{\mathrm{c},R}(\beta),c\,\beta) \, ,
\end{align}
where $c$ is a fixed prefactor.
From Eqs.~(\ref{Eq:Rsc}) and (\ref{Eq:Rcross}) one can show that
for $\beta\to\infty$ the pseudocritical coupling converges to $\alphac$ as
\begin{align}
\label{Eq:fitfunction}
\alpha_{\mathrm{c},R}(\beta)
	=
	\alphac + A \, \beta^{-e} \, ,
	\qquad
	e=1/\nu + \omega \, ,
\end{align}
where $A$ is a non-universal constant \cite{FSS14, toldin15}.

For our numerical analysis, we consider the correlation length $\xi_x$ along the imaginary-time
axis and define the RG-invariant $R_\xi$ as
\begin{align}
\label{Eq:Rcor}
R_\xi \equiv \xi_x / \beta
	= \frac{1}{2\pi} \sqrt{\frac{\chi_x(\im \Omega_0)}{\chi_x(\im \Omega_1)}-1} \, .
\end{align}
In analogy to the definition of the spatial correlation length \cite{Sandvik10},%
\footnote{%
In lattice simulations, it is standard practice to extract the spatial
correlation length
$
\xi_\mathrm{space} = \frac{1}{\delta q} \sqrt{\frac{C(Q)}{C(Q+\delta Q)}-1}
$
from
the two-point correlation function $C(q)$ in momentum space \cite{Sandvik10},
which is evaluated at the ordering vector $Q$ and the closest
momentum $Q+\delta q$, where $\delta q = 2\pi / L$ is the resolution in momentum space.
The shift $\delta q$ takes the long-wavelength fluctuations near the ordering
vector into account. Note that there is not a unique way to define a correlation length on a finite lattice, but this
is a common one.}
we identify the inverse
temperature  $\beta$ with the system size in imaginary-time direction and the Matsubara
frequency $\Omega_0 = 0$ with the ordering vector;
we use $\Omega_1 = 2\pi / \beta$ as the resolution in frequency space.
The lowest Matsubara frequencies capture the long-time decay of the spin autocorrelations, as shown in Sec.~\ref{Sec:SusDetails}.
We can derive a similar RG-invariant observable from Eq.~(\ref{Eq:Rcor}),
\begin{align}
\label{Eq:Rcor2}
\tilde{R}_\xi = \frac{\chi_x(\im \Omega_1)}{\chi_x(\im \Omega_0)} \, ,
\end{align}
which obviously has the same properties near criticality.
The definition of $\tilde{R}_\xi$ stems from the correlation ratio, another common estimator that is used
in FSS analyses.

For the SU(2)-symmetric spin-boson model, the perturbative RG has predicted a critical phase \cite{sengupta,SBV99}
in which the correlation length $\xi$ scales with the system size $\beta$. We have presented
numerical evidence that $R_\xi = \xi_x/\beta$ is indeed finite in the CR phase. The existence
of (un)stable fixed points requires that $R_\xi$ is scale-invariant at these points
and that they can be analyzed using a FSS scheme.
In our main article, we have identified the two scale-invariant points with the stable fixed point of the CR phase at $\alpha_\mathrm{CR}$
and the unstable fixed point at the quantum critical coupling $\alpha_\mathrm{QC}$.
In the following, we show how we determined $\alpha_\mathrm{CR}$ and $\alpha_\mathrm{QC}$
as well as the critical exponents at $\alpha_\mathrm{QC}$.

\subsection{Low-frequency behavior of the susceptibility\label{Sec:SusDetails}}

\begin{figure}[ht!]
\includegraphics[width=\columnwidth]{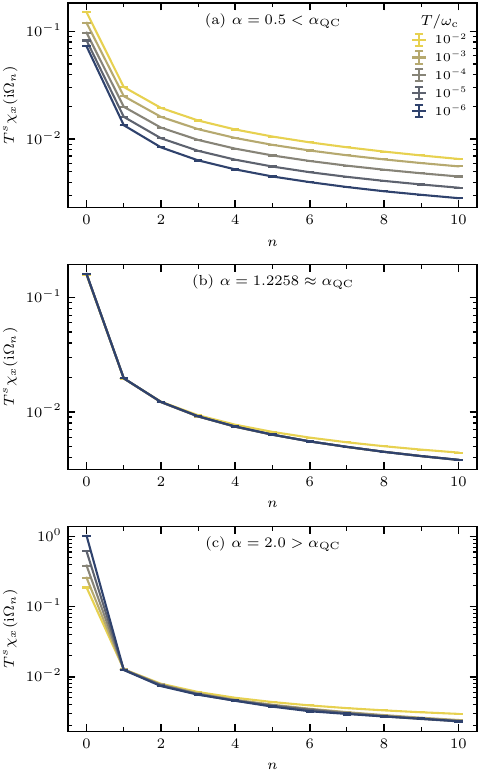}
\caption{%
Rescaled dynamical spin susceptibility $T^s \chi_x(\im \Omega_n)$ as a function of the Matsubara index $n$ for
different temperatures (a) in the CR phase, (b) at the critical coupling, and (c) in the L phase. Here, $s=0.75$.
}
\label{fig:orderp}
\end{figure}

Before we perform our FSS analysis of the correlation length $\xi_x$,
we discuss the low-frequency behavior of the dynamical spin
susceptibility, as it enters the definition of $\xi_x$ in Eq.~\eqref{Eq:Rcor}.
We first recapitulate the long-time decay of the correlation function $\chi_x(\tau)=\langle S^x(\tau) S^x(0)\rangle$:
in the CR phase, $\chi_x(\tau) \sim 1/\tau^{1-s}$, as predicted by the perturbative RG \cite{sengupta,SBV99,rg_bfk},
whereas in the long-range-ordered L phase $\chi_x(\tau) \sim m^2_\mathrm{loc}$.
We can now transform the dynamical susceptibility from imaginary time $\tau$ to Matsubara frequencies $\Omega_n = 2\pi n / \beta$ via
its Fourier transformation
\begin{align}
\chi_x(\im\Omega_n)
	=
	\int_0^{\beta} d\tau \, e^{\im \Omega_n \tau} \left\langle S^x(\tau) S^x(0) \right\rangle \, .
\end{align}
Then, the static susceptibility $\chi_x \equiv \chi_x(\im\Omega_0)$ diverges in both phases for $T\to0$:
we obtain $\chi_x(T) \propto T^{-s}$ in the CR phase, whereas $\chi_x(T) = m^2_\mathrm{loc}/T$ in the L phase.
At low but finite Matsubara frequencies, the power-law decay of $\chi_x(\tau)$ in the CR phase leads to $\chi_x(\im\Omega_n) \propto \left| \im \Omega_n \right|^{-s}$. While $\chi_x(\im\Omega_n)$ always remains finite at a given frequency $\im\Omega_n$,
fixing the Matsubara index $n$---\eg, considering the first nonzero component $\chi_x(\im\Omega_1)$, which enters the definition of the correlation length in 
Eq.~\eqref{Eq:Rcor}---leads to a divergent behavior of $\chi_x(\im\Omega_n) \propto T^{-s}$ if we perform our FSS analysis in inverse temperature.
As a result, the ratio $\chi_x(\im \Omega_0)/\chi_x(\im \Omega_1)$ remains finite and leads to $\xi_x \sim \beta$, as it
is expected within the critical phase. This behavior is also confirmed by our QMC simulations: In Fig.~\ref{fig:orderp}
we show the rescaled susceptibility $T^s \chi_x(\im \Omega_n)$ as a function of the Matsubara index $n$
for different temperatures. We observe that within the CR phase the ratio of $\chi_x(\im \Omega_0)$ and $\chi_x(\im \Omega_1)$
in Fig.~\ref{fig:orderp}(a) remains finite, whereas each element still experiences subleading corrections. Only at the two nontrivial fixed points, \ie, at the critical
coupling $\alpha_\mathrm{QC}$ shown in Fig.~\ref{fig:orderp}(b) and at the stable fixed-point coupling $\alpha_\mathrm{CR}$ (not shown),
subleading corrections vanish, such that $T^s \chi_x(\im\Omega_n)$ reaches a constant for all temperatures $T/\omega_\mathrm{c} \lesssim 10^{-3}$;
therefore $\chi_x(\im\Omega_n) \propto \left| \im \Omega_n\right|^{-s}$ holds over a wide range of frequencies.
Even within the ordered phase, the finite-frequency components fulfill $\chi_x(\im\Omega_n) \propto \left| \im \Omega_n \right|^{-s}$, as shown in Fig.~\ref{fig:orderp}(c) and first observed in Ref.~\cite{otsuki13}. However, the static component diverges as $1/T$, such that the correlation length $\xi_x/\beta$ in 
Eq.~\eqref{Eq:Rcor} diverges with decreasing temperature.

\subsection{Estimation of the critical exponent $1/\nu$\label{Sec:EstimateExp}}

We first want to explain how we determined the critical exponent $1/\nu$
shown in our main paper.
To benchmark our analysis, we consider the U(1)-symmetric two-bath spin-boson model (also dubbed XY-symmetric Bose-Kondo model), \ie, $\alpha_z = 0$, for which $1/\nu$ had been determined using a variational MPS approach \cite{bruo14}. For our subsequent analysis, we use the QMC results recently presented in Ref.~\onlinecite{weber21}; a detailed FSS analysis for $1/\nu$ remained left for this paper.
Afterwards, we will apply the same procedure to the SU(2)-symmetric case.
Our FSS analysis follows the strategy described in Ref.~\onlinecite{toldin15}.

To determine the critical exponent $1/\nu$ from the scaling form in Eq.~(\ref{Eq:Rsc}),
we need to fit our QMC data $R(\alpha,\beta)$ to Eq.~(\ref{Eq:Rsc}).
However, even in the absence of subleading corrections, the universal scaling
function $f_R(w)$ is not known, so that we approximate it by a Taylor expansion around criticality to obtain
\begin{align}
\label{Eq:R_fit}
R = R^\ast
+ \sum_{n=1}^{\nmax} a_n \left( \alpha - \alphac \right)^n \beta^{n/\nu} \, .
\end{align}
Then, we can use $R^\ast$, $1/\nu$, and $\{a_n\}$ as free parameters for a $\chi^2$ fit
of our QMC data to Eq.~(\ref{Eq:R_fit}). The accuracy of the fit function depends on
the maximum expansion order $\nmax$. We restrict our input data to a small
interval of $\alpha$ around $\alphac$.

\begin{table}[t]
\caption{\label{Tab:1}%
Benchmark of our finite-size-scaling analysis for the U(1)-symmetric spin-boson model at $s=0.8$.
Results correspond to different fits of $\tilde{R}_\xi$ to Eq.~(\ref{Eq:R_fit}) for $\alpha \in [0.75,0.78]$.
$\bmin$ is the minimum inverse temperature taken into account for the fits,
$\beta_\mathrm{max} \, \omega_\mathrm{c}=10^6$,
and $\chi^2/\mathrm{DOF}$ is the normalized goodness-of-fit measure.
The optimal fit 
 is highlighted in red;
the extracted critical exponent is in excellent agreement with the
MPS result $1/\nu =0.106(3)$ \cite{bruo14,bruoThesis}.
As input data we use the QMC results recently presented in Ref.~\onlinecite{weber21}.
}
\begin{ruledtabular}
\begin{tabular}{ccccc}
			 & $\bmin \, \omega_\mathrm{c}$&	$\alpha_\mathrm{QC}$		&	$1/\nu$		&	$\chi^2/\mathrm{DOF}$  \\
 \hline
			& $10^3$	&	0.7618(18)\phantom{0}	&	0.078(13)\phantom{0}		&	21.1		\\
$\nmax=1$	& $10^4$	&	0.7622(3)\phantom{00}		&	0.107(2)\phantom{00}		&	\phantom{0}1.6		\\
			& $10^5$	& 	0.7626(7)\phantom{00}		&	0.112(8)\phantom{00}		&	\phantom{0}1.6		\\
\hline
			& $10^3$	&	0.7607(17)\phantom{0}	& 	0.076(12)\phantom{0}		&	19.0		\\
$\nmax=2$	& $10^4$	& 	\textcolor{red}{0.76237(17)}	& 	\textcolor{red}{0.1073(15)}	&	\phantom{00}\textcolor{red}{1.04}		\\
			& $10^5$	& 	0.7628(6)\phantom{00}		&	0.111(5)\phantom{00}		&	\phantom{00}1.23		\\
\hline
			& $10^3$	&	0.7596(17)\phantom{0}	&	0.063(9)\phantom{00}		&	16.8		\\
$\nmax=3$	& $10^4$	& 	0.76236(17) 	&	0.1064(17)	&	\phantom{00}1.04		\\
			& $10^5$	&	0.7628(6)\phantom{00}		&	0.108(6)\phantom{00}		&	\phantom{00}1.22 		\\
\end{tabular}
\end{ruledtabular}
\end{table}
Benchmark results for the U(1)-symmetric
spin-boson model at $s=0.8$ are summarized in Table~\ref{Tab:1}.
We repeated the fits for $\nmax\in \{1,2,3\}$ and also varied the lower bound $\bmin$ of
inverse temperatures taken into account for the fit.
One measure to assess the quality of the fit is the
goodness-of-fit measure $\chi^2/\mathrm{DOF}$ which
is normalized by the total number of degrees of freedom.
A good fit is obtained if each data point varies on average by one standard deviation
from the fit function; this corresponds to $\chi^2/\mathrm{DOF} \approx 1$.
Table~\ref{Tab:1} reveals that we have to
choose $\bmin \, \omega_\mathrm{c} \gtrsim 10^4$ in order to obtain good fits. While $\nmax = 1$
already gives decent results, optimal fits require $\nmax \geq 2$.
However, an increasing number of free fit parameters will also increase the uncertainty of our estimates.
As an optimal choice to weigh between complexity of the fit ansatz and accuracy, we choose $\nmax = 2$ and
$\bmin \, \omega_\mathrm{c} = 10^4$. With these parameters, we estimate $1/\nu = 0.1073(15)$, which is in
excellent agreement with the MPS result of $1/\nu =0.106(3)$ \cite{bruo14,bruoThesis}.

\begin{figure}[tb]
\includegraphics[width=\columnwidth]{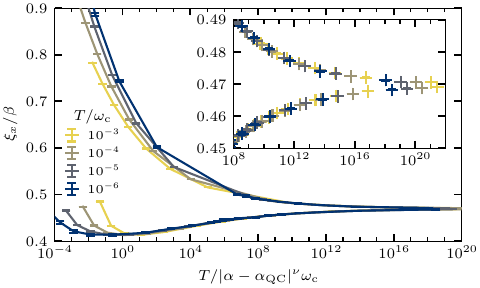}
\caption{%
Data collapse for $\xi_x / \beta$ near the QC fixed point of the U(1)-symmetric
spin-boson model at $s=0.8$. The inset shows
a detailed view of the critical region. We used $\alpha_\mathrm{QC}$ and $\nu$
as determined from the FSS analysis in Table~\ref{Tab:1}.
}
\label{fig:collapseU1}
\end{figure}

To obtain a precise and reliable estimate for the critical exponent $1/\nu$,
it is important to check the stability of our fits.
Even if $\chi^2/ \mathrm{DOF} \approx 1$, our results could still suffer from over- and underfitting
in distinct parameter regimes. To get another measure for the convergence
of our fits, we systematically vary $\nmax$ and $\bmin$ in Table~\ref{Tab:1}.
We observe that our optimal result at $\nmax=2$ and $\bmin \, \omega_\mathrm{c} = 10^4$
does not change anymore within error bars if we increase either of the parameters.
This is a clear signature that our fits have converged. We have also checked that
the inclusion of subleading corrections to the fit ansatz in Eq.~(\ref{Eq:R_fit}) does not affect our results.
To this end, we included a Taylor expansion of $g_R(w)$ in Eq.~(\ref{Eq:R_fit}), so that the expansion coefficients as well as the exponent $\omega$
of the subleading corrections become additional fit parameters.
Note that the convergence of the $\chi^2$ fits becomes increasingly difficult
with an increasing number of free parameters, therefore we restrict our analysis to the strategy presented
in Table~\ref{Tab:1}. As our QMC method reaches very low temperatures and our QMC estimate
is in excellent agreement with a previous MPS result, systematic errors of $1/\nu$
originating from subleading corrections are expected to be at most of the order of our error bars.
Eventually, the quality of our estimate for $1/\nu$ is also confirmed by the excellent data collapse
which we present in Fig.~\ref{fig:collapseU1} for the U(1)-symmetric spin-boson model.

\begin{figure*}[p!]
\includegraphics[width=\textwidth]{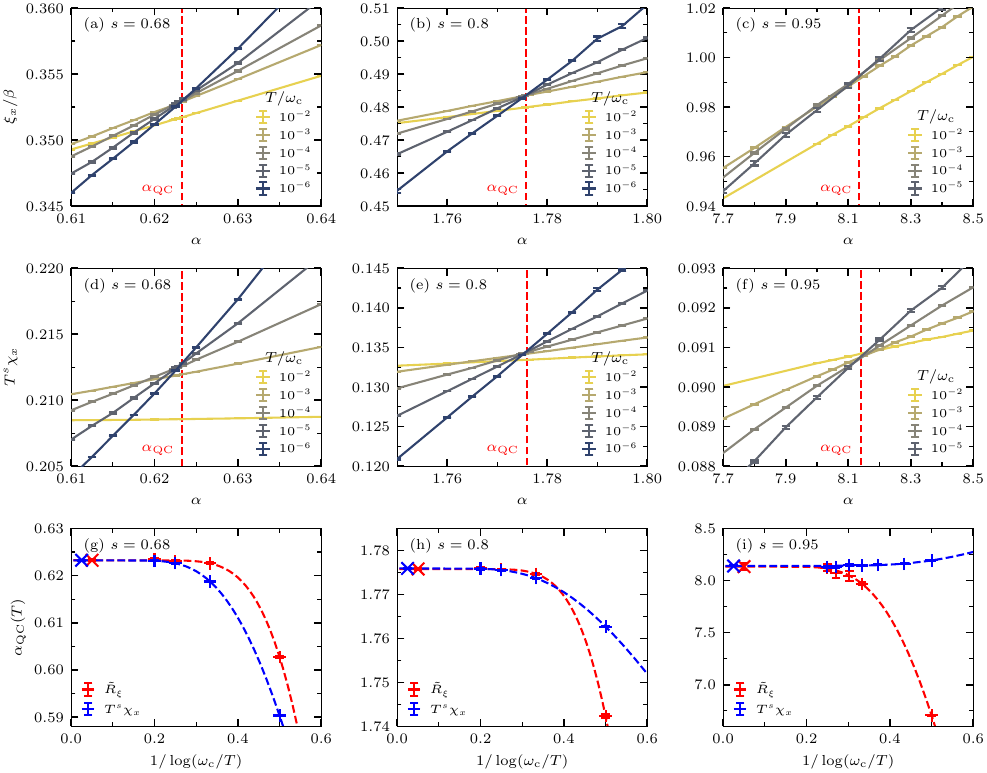}
\caption{%
Estimation of the quantum critical coupling $\alpha_\mathrm{QC}$ via the crossing method for different bath exponents $s$.
We show the crossings of (a)--(c) the correlation length $\xi_x / \beta$ and (d)--(f) the rescaled susceptibility $T^s \chi_x$
close to $\alpha_\mathrm{QC}$ for different temperatures. We extract the pseudocritical coupling $\alpha_\mathrm{QC}(T)$
from the crossings between data pairs $(T, T/c)$ with $c=10$ and plot the results in panels (g)--(i). Using power-law fits
as defined in Eq.~(\ref{Eq:fitfunction}) (shown as dashed lines), we extrapolate $\alpha_\mathrm{QC}=\alpha_\mathrm{QC}(T\to 0)$
and include the fitted values near $T=0$. All our estimates for $\alpha_\mathrm{QC}$ are collected in Table~\ref{Tab:2}.
}
\label{fig:crossingsQC}

\captionof{table}{\label{Tab:2}%
Collection of the fixed-point couplings $\alpha_\mathrm{CR}$ and $\alpha_\mathrm{QC}$ as well as the inverse correlation-length exponent, $1/\nu$,
of the SU(2)-symmetric spin-boson model for different bath exponents $s$.
We compare different approaches to extract $\alpha_\mathrm{CR}$ and $\alpha_\mathrm{QC}$ from FSS:
For both couplings, we performed a crossing analysis of $T^s \chi_x$ and $\tilde{R}_\xi$.
Moreover, we estimated $\alpha_\mathrm{QC}$ and $1/\nu$ from fits to the scaling form (\ref{Eq:R_fit}).
All our estimates agree well within error bars. For $\alpha_\mathrm{CR}$ and $\alpha_\mathrm{QC}$,
best precision is obtained from $T^s \chi_x$.
The fact that $\alpha_\mathrm{QC}$ from Eq.~(\ref{Eq:R_fit}) agrees well with the other estimates for $\alpha_\mathrm{QC}$ indicates
that subleading corrections are small for the estimation of $1/\nu$ via $\tilde{R}_\xi$.
}
\begin{ruledtabular}
\begin{tabular}{c|cc|ccc|c}
s &
$\alpha_\mathrm{CR}$ from $T^s\chi_x$ &
$\alpha_\mathrm{CR}$ from $\tilde{R}_\xi$ &
$\alpha_\mathrm{QC}$ from $T^s\chi_x$	&
$\alpha_\mathrm{QC}$ from $\tilde{R}_\xi$ &
$\alpha_\mathrm{QC}$ from Eq.~(\ref{Eq:R_fit}) &
$1/\nu$
 \\
 \hline
0.655 & 0.2730(11)\phantom{0} & -- & 0.3597(9)\phantom{00} & -- & 0.3590(19)\phantom{0}& 0.025(4) \\
0.66\phantom{0} & 0.2267(14)\phantom{0} & -- & 0.4353(3)\phantom{00} & 0.4368(9)\phantom{00} & 0.4348(8)\phantom{00} & 0.052(3) \\
0.67\phantom{0} & 0.1856(4)\phantom{00} & -- & 0.53589(16) & 0.5364(4)\phantom{00} & 0.5358(3)\phantom{00} & 0.101(4) \\
0.68\phantom{0} & 0.1609(5)\phantom{00} & 0.162(2)\phantom{000} & 0.62325(7)\phantom{0} & 0.62326(19) & 0.62300(19) & 0.118(3) \\
0.7\phantom{00} & 0.1287(15)\phantom{0} & 0.1296(14)\phantom{0} & 0.78792(10) & 0.7879(2)\phantom{00} & 0.78786(17) & 0.151(2) \\
0.75\phantom{0} & 0.0861(2)\phantom{00} & 0.0871(5)\phantom{00} & 1.22554(10) & 1.2259(2)\phantom{00} & 1.22539(14) & 0.192(2) \\
0.8\phantom{00} & 0.0610(2)\phantom{00} & 0.0600(10)\phantom{0} & 1.77592(13) & 1.77579(19) & 1.7754(2)\phantom{00} & 0.195(3) \\
0.85\phantom{0} & 0.04228(10) & 0.0424(3)\phantom{00} & 2.5703(4)\phantom{00} & 2.5707(10)\phantom{0} & 2.5701(6)\phantom{00} & 0.163(2) \\
0.9\phantom{00} & 0.02675(5)\phantom{0} & 0.0267(2)\phantom{00} & 4.0130(8)\phantom{00} & 4.010(6)\phantom{000} & 4.013(3)\phantom{000} & 0.104(4) \\
0.95\phantom{0} & 0.01286(2)\phantom{0} & 0.01309(10) & 8.140(4)\phantom{000} & 8.13(4)\phantom{0000} & 8.113(14)\phantom{00} & 0.053(3) \\
0.975 & 0.006336(6) & -- & 16.32(4)\phantom{00000} & -- & -- & -- \\

\end{tabular}
\end{ruledtabular}

\end{figure*}

Having benchmarked our FSS analysis for the U(1)-symmetric spin-boson model, we applied the
same strategy to the SU(2)-symmetric model. We made sure that our optimal fits always contained
at least three data sets covering a temperature range
with $\beta_\mathrm{max} / \bmin$ between 10 and 100. Our final results are shown in
our main paper, but also collected in Table~\ref{Tab:2}. For the SU(2)-symmetric model,
a data collapse is also shown in the main paper. We want to emphasize that
our precise estimation of $1/\nu$ was only possible due to the development of the
efficient wormhole QMC method presented in Ref.~\onlinecite{weber21}.

\subsection{Determination of the fixed-point couplings}

\subsubsection{Quantum critical coupling $\alpha_\mathrm{QC}$}

We have obtained a first estimate of the quantum critical coupling $\alpha_\mathrm{QC}$
during the estimation of the critical exponent $1/\nu$ using the fitting procedure in Eq.~(\ref{Eq:R_fit}).
In the following, we want to apply the crossing method described above to get alternative estimates for $\alpha_\mathrm{QC}$.
The crossing method has the advantage that a FSS scaling of the pseudocritical coupling
via Eq.~(\ref{Eq:fitfunction}) also takes into account subleading corrections. We apply the crossing method
to the correlation length $\xi_x / \beta$ defined in Eq.~(\ref{Eq:Rcor})
[equivalently, we also use the correlation ratio $\tilde{R}_\xi$ in Eq.~(\ref{Eq:Rcor2})]
and to the rescaled susceptibility $T^s \chi_x$. Eventually, we will compare the quality of all our estimates,
which are collected in Table~\ref{Tab:2}.

Figure~\ref{fig:crossingsQC} shows selected results of our FSS analysis of $\xi_x/\beta$ and $T^s \chi_x$
around $\alpha_\mathrm{QC}$ for bath exponents $s$ chosen over the entire relevant parameter range for which
we observe quantum critical behavior. For each exponent $s$, we show the crossings of $\xi_x / \beta$
in Figs.~\ref{fig:crossingsQC}(a)--(c) and $T^s \chi_x$ in Figs.~\ref{fig:crossingsQC}(d)--(f) as a function of
$\alpha$ and for different temperatures $T/\omega_\mathrm{c}$. For both observables, the pseudocritical couplings
$\alpha_\mathrm{QC}(T)$ are extracted from the crossings of data sets at temperatures $T$ and $T/c$ with $c=10$, as defined in Eq.~(\ref{Eq:Rcross}), and shown in Figs.~\ref{fig:crossingsQC}(g)--(i) as a function of temperature. For bath exponents $s\gtrsim s^\ast$, \ie, for $s=0.68$ and still for $s=0.8$, we observe that the extracted crossings from $\xi_x / \beta$ converge faster to $\alpha_\mathrm{QC}$ than the ones extracted from $T^s \chi_x$; this is somehow expected because $\xi_x / \beta$ is an improved estimator.
However, for $s\lesssim 1$, \ie, for $s=0.95$, the pseudocritical coupling extracted from $T^s \chi_x$ converges much faster. Note that we use a bootstrap analysis to get reliable estimates for the error bars of the pseudocritical couplings; for each bootstrap sample we use a quadratic fit to determine the crossing. Finally, to estimate $\alpha_\mathrm{QC}$ we fit the extracted $\alpha_\mathrm{QC}(T)$ shown in Figs.~\ref{fig:crossingsQC}(g)--(i) to Eq.~(\ref{Eq:fitfunction}) using the Marquardt-Levenberg method implemented in Gnuplot. The fits are shown as dashed lines in Figs.~\ref{fig:crossingsQC}(g)--(i) and the fitted $\alpha_\mathrm{QC}$ are included in our plots close to $T=0$. We observe that all our fits for $\xi_x / \beta$ and $T^s \chi_x$ are in excellent agreement with each other.

For a quantitative comparison of our estimates, we list all our results in Table~\ref{Tab:2}.
First of all, we want to emphasize that the two estimates from the crossing analysis and
the estimate from extracting $1/\nu$ are in excellent agreement within error bars. Although
we did not include subleading corrections in our determination of $1/\nu$ using Eq.~(\ref{Eq:R_fit}),
the extracted $\alpha_\mathrm{QC}$ values are in good agreement with the others. In particular,
its error bars are of similar size as for the estimates from the crossing analysis of $\tilde{R}_\xi$.
However, we find that our estimates from the crossings of $T^s \chi_x$ are significantly more
precise than the other estimates. This is a direct consequence of the fact that results for $T^s \chi_x$
in Figs.~\ref{fig:crossingsQC}(d)--(f) have smaller error bars than the results for $\xi_x/\beta$
in Figs.~\ref{fig:crossingsQC}(a)--(c). Indeed, the estimator for $\xi_x/\beta$ in Eq.~(\ref{Eq:Rcor})
includes a ratio of two expectation values, which is known to enhance statistical fluctuations
in the final estimator. Because of this, we were not able to extract reliable crossings for $s=0.655$
and $s=0.975$, whereas the crossings of $T^s \chi_x$ remain clearly visible for $s\to s^\ast$
and $s\to1$.

All in all, because the crossing analysis of $T^s \chi_x$ gives the most precise estimates for $\alpha_\mathrm{QC}$, we use these results in our main paper. However, we were not able to extract similarly reliable critical exponents $1/\nu$ from the susceptibility. Using an analysis as in Sec.~\ref{Sec:EstimateExp}, we were neither able to accurately reproduce the benchmark result for the U(1)-symmetric spin-boson model nor our results for the SU(2)-symmetric case, but instead observed a slow drift of the estimated $1/\nu$ depending on the temperature range used for the scaling analysis. This leads us to conclude that scaling corrections are much larger for the susceptibility than for the correlation length.

\subsubsection{Stable fixed-point coupling $\alpha_\mathrm{CR}$}

\begin{figure*}[p!]
\includegraphics[width=\textwidth]{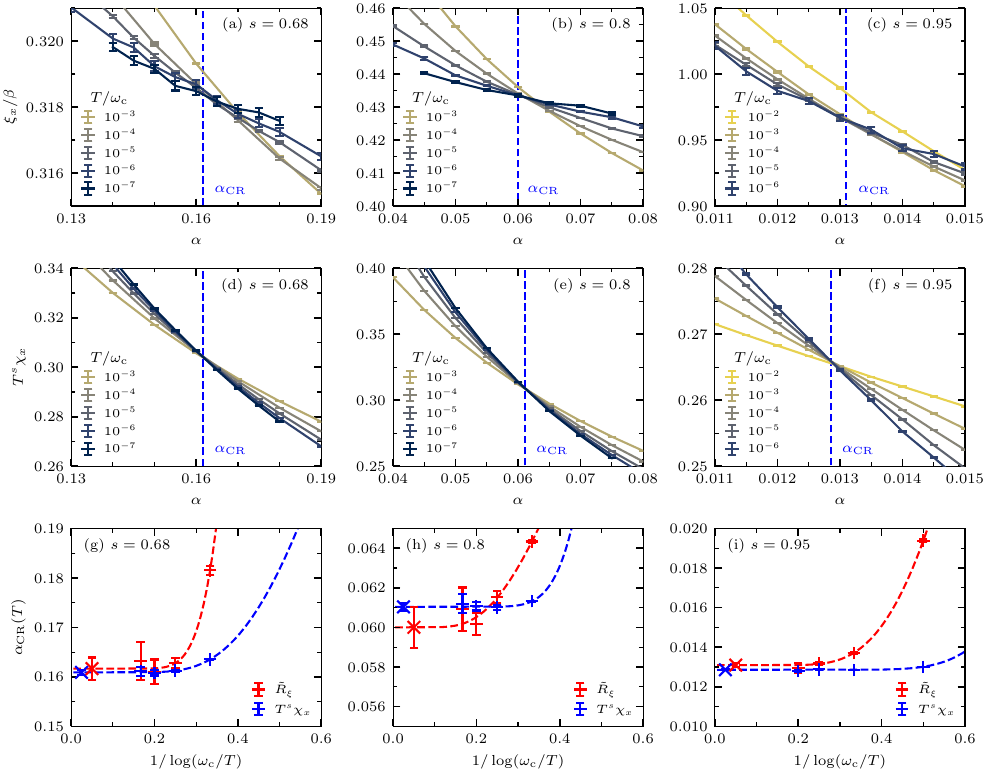}
\caption{%
Estimation of the coupling $\alpha_\mathrm{CR}$ for the stable fixed point via the crossing method for different bath exponents $s$.
We show the crossings of (a)--(c) the correlation length $\xi_x / \beta$ and (d)--(f) the rescaled susceptibility $T^s \chi_x$
close to $\alpha_\mathrm{CR}$ for different temperatures. We extract the pseudocritical coupling $\alpha_\mathrm{CR}(T)$
from the crossings between data pairs $(T, T/c)$ with $c=10$ and plot the results in panels (g)--(i). Using power-law fits
as defined in Eq.~(\ref{Eq:fitfunction}) (shown as dashed lines), we extrapolate $\alpha_\mathrm{CR}=\alpha_\mathrm{CR}(T\to 0)$
and include the fitted values near $T=0$. All our estimates for $\alpha_\mathrm{CR}$ are collected in Table~\ref{Tab:2}.
}
\label{fig:crossingsCR}
\vspace{0.5cm}
\includegraphics[width=\textwidth]{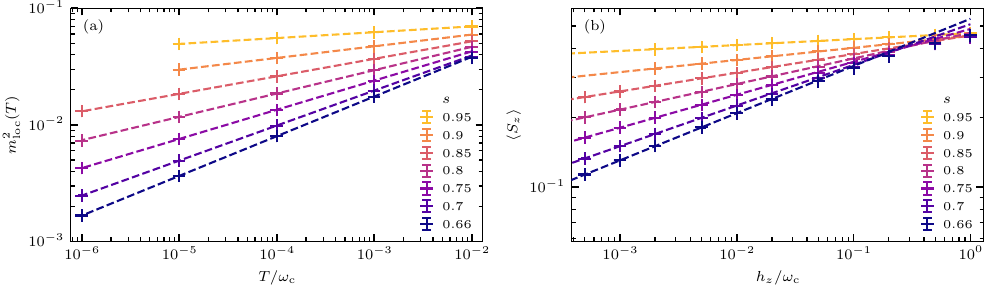}
\caption{%
Confirmation of the hyperscaling relations at the critical coupling $\alpha_\mathrm{QC}$.
(a) Spin-spin correlation function at $\tau=\beta/2$, \ie, $m^2_\mathrm{loc}(T)$ defined in Eq.~(\ref{Eq:orderp}), as a function of temperature for
different bath exponents $s$. Dashed lines represent
power-law fits to $a \, T^{2\beta' / \nu}$
where the exponent is fixed by the hyperscaling relation in Eq.~(\ref{eq:beta}) and $a$ is the only free fitting parameter.
(b)~Magnetization $\langle S^z \rangle$ as a function of the applied field $h_z$ for different $s$.
We only show results converged in temperature.
Power-law fits to $a \, h_z^{1/\delta}$
with $\delta$ given by Eq.~(\ref{eq:delta}) are shown as dashed lines.
}
\label{fig:hyper}
\end{figure*}

We have seen in our main paper that $\xi_x/\beta$ shows an additional crossing at the stable fixed point of the quantum critical phase. We can also apply a crossing analysis, as described above, to extract the corresponding fixed-point coupling $\alpha_\mathrm{CR}$. For selected bath exponents $s$, results are shown in Fig.~\ref{fig:crossingsCR}, whereas all our results are included in Table~\ref{Tab:2}. As it has already been the case for $\alpha_\mathrm{QC}$, we get significantly more precise estimates from the crossings of $T^s \chi_x$. Therefore, we have run our simulations
only long enough to get good convergence for $T^s \chi_x$; note that we could have easily obtained results with smaller error bars for $\xi_x/\beta$ as well, because our QMC simulations are very efficient in the weak-coupling regime.
For both observables, our estimated $\alpha_\mathrm{CR}$ are consistent within error bars. For $s$ very close to $s^\ast$, error bars for $\xi_x/\beta$ were too large to get reliable estimates; therefore, we did not include them in Table~\ref{Tab:2}. In our main paper, we included the fixed-point couplings $\alpha_\mathrm{CR}$ as determined from the crossings of $T^s \chi_x$.

We note that it is conceptually non-trivial to determine the location of a stable intermediate-coupling fixed point in a numerical simulation of a microscopic model, and we are not aware of previous work in this direction. In the present model, choosing $\alpha=\alpha_\mathrm{CR}$ in $\mathcal{H}$ [\cf Eq.~(1)] implies that the RG flow of $\alpha$ is stationary, i.e., the prefactor of the leading irrelevant perturbation to the CR fixed point---which is essentially $(\alpha-\alpha_\mathrm{CR})$---vanishes. This then implies scale invariance, as observed numerically. In more complicated models, RG flow and fixed points are embedded in a higher-dimensional space of effective couplings which is not fully accessible by varying microscopic couplings, but finding a point where the prefactor to the leading irrelevant perturbation vanishes should be possible.

\section{Hyperscaling relations for the magnetization exponents\label{Sec:hyperhyper}}

Apart from the correlation length exponent $\nu$, we can define additional
critical exponents at the quantum phase transition. Here,
we use the local moment $m_\mathrm{loc}$ as a zero-temperature order parameter.
For the SU(2)-symmetric spin-boson model, we have
\begin{align}
\label{eq:sc_beta}
m_\mathrm{loc}(\alpha \gtrsim \alpha_\mathrm{QC})
	\propto
	(\alpha - \alpha_\mathrm{QC})^{\beta'}
\end{align}
when entering the localized phase. The critical exponent $\beta'$ also applies
under variation of the bath exponent $s$. If we apply an additional magnetic
field $h_i S^i$ to the Hamiltonian, we can measure the response of the system at the critical point
$\alpha=\alpha_\mathrm{QC}$ via
\begin{align}
\label{eq:sc_delta}
m_\mathrm{loc}(h_i; \alpha = \alpha_\mathrm{QC})
	\propto
	h_i^{1/\delta} \, .
\end{align}
The magnetic field reduces the SU(2) symmetry to U(1).
If we assume hyperscaling, the critical exponents become
\begin{align}
\label{eq:beta}
\frac{\beta'}{\nu} &= \frac{1-s}{2} \, ,
	\\
\delta &= \frac{1+s}{1-s} \, .
\label{eq:delta}
\end{align}
For further details on the definition of critical exponents and hyperscaling for spin-boson models, see Ref.~\onlinecite{bruo14}.

\begin{figure}[t]
\includegraphics[width=\columnwidth]{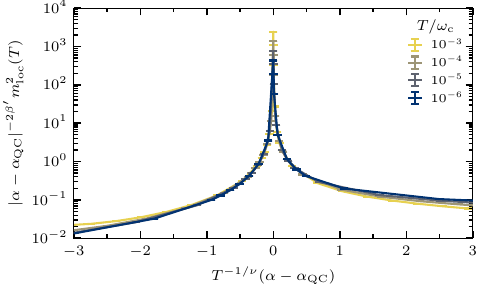}
\caption{%
Data collapse for $m^2_\mathrm{loc}(T)$ near the QC fixed point
using the scaling ansatz of Eq.~(\ref{Eq:scaling_ansatz_mloc}).
We used $s=0.75$, $\alpha_\mathrm{QC}$ and $\nu$
as given in Table~\ref{Tab:2}, and $\beta'$ determined by the hyperscaling
relation in Eq.~(\ref{eq:beta}).
}
\label{fig:collapse_mloc}
\end{figure}

\subsection{Numerical verification of hyperscaling}

In the following, we show that the exponents $\beta'$ and $\delta$ indeed fulfill the hyperscaling relations.
It is difficult to determine $\beta'$ directly from Eq.~(\ref{eq:sc_beta}) because even for the
lowest temperatures reached in our QMC simulations, $m_\mathrm{loc}$ has not yet converged
close to the critical point (see, \eg, Fig.~\ref{fig:localmom}).
However, at $\alpha=\alpha_\mathrm{QC}$ the order parameter defined in Eq.~(\ref{Eq:orderp}) fulfills the finite-temperature
scaling
\begin{align}
\label{eq:betanu_fit}
m^2_\mathrm{loc}(T)
	\propto
	T^{2\beta'/\nu} \, ,
\end{align}
which gives direct access to the ratio $\beta'/\nu$.
Figure~\ref{fig:hyper}(a) shows the temperature dependence of the order parameter for different bath exponents $s$. Power-law fits to Eq.~(\ref{eq:betanu_fit}) with exponents determined by Eq.~(\ref{eq:beta}) are in excellent
agreement with our numerical data. To determine the exponent $\delta$, we apply a magnetic field
in $z$ direction. Then, the local moment can be measured easily from our QMC simulations
via $\langle S^z \rangle$. In Fig.~\ref{fig:hyper}(b), we show $\langle S^z \rangle$ as a function of $h_z$
for different $s$. Again, power-law fits to Eq.~(\ref{eq:sc_delta}) with exponents given by Eq.~(\ref{eq:delta}) are in excellent
agreement with our data. Hence, we conclude that hyperscaling is fulfilled for the SU(2)-symmetric
spin-boson model. Therefore, we have calculated $\beta'$ via Eq.~(\ref{eq:beta}) from $\nu$, as presented in our main paper.

\subsection{Data collapse of the local moment $m^2_\mathrm{loc}(T)$}

Further confirmation of the hyperscaling relations can be obtained from
a data collapse for the local moment.
For finite temperatures, $m^2_\mathrm{loc}(T)$ follows the scaling form
\begin{align}
\label{Eq:scaling_ansatz_mloc}
m^2_\mathrm{loc}(T)
	=
	\left| \alpha - \alpha_\mathrm{QC} \right|^{2 \beta'} \tilde{f}(\beta^{1/\nu} \left(\alpha - \alpha_\mathrm{QC}\right)) \, ,
\end{align}
where $\tilde{f}$ is a universal scaling function, with the limiting behavior $\tilde{f}(x\to\infty)\to\rm const$ [\cf Eq.~\eqref{eq:sc_beta}] and $\tilde{f}(x\to-\infty)\to0$. Figure~\ref{fig:collapse_mloc} depicts the corresponding data collapse
for $s=0.75$. Here, the exponent $\beta'$ has been fixed by the hyperscaling relation in Eq.~(\ref{eq:beta}).
We find an excellent data collapse which is further support for the validity hyperscaling.

%
%
%
%


\begin{thebibliography}{99}

\bibitem{kaplan09}
D. B. Kaplan, J.-W. Lee, D. T. Son, and M. A. Stephanov,
Phys Rev. D \textbf{80}, 125005 (2009).

\bibitem{halperin74}
B. I. Halperin, T. C. Lubensky, and S.-k. Ma,
Phys. Rev. Lett. \textbf{32}, 292 (1974).

\bibitem{braun14}
J. Braun, H. Gies, L. Janssen, and D. Roscher,
Phys. Rev. D \textbf{90}, 036002 (2014).

\bibitem{gies06}
H. Gies and J. Jaeckel,
Eur. Phys. J. C \textbf{46}, 433 (2006).

\bibitem{Nienhuis79}
B. Nienhuis, A. N. Berker, E. K. Riedel, and M. Schick,
Phys. Rev. Lett. \textbf{43}, 737 (1979).

\bibitem{nahum15}
A. Nahum, J. T. Chalker, P. Serna, M. Ortu\~no, and A. M. Somoza,
Phys. Rev. X \textbf{5}, 041048 (2015).

\bibitem{wang17}
C. Wang, A. Nahum, M. A. Metlitski, C. Xu, and T. Senthil,
Phys. Rev. X \textbf{7}, 031051 (2017).

\bibitem{Nahum20}
A. Nahum,
Phys. Rev. B \textbf{102}, 201116(R) (2020).

\bibitem{leggett}
A.~J. Leggett {\em et~al.},
Rev. Mod. Phys. {\bf 59}, 1 (1987).

\bibitem{SBV99}
S. Sachdev, C.~Buragohain, and M. Vojta,
Science {\bf 286}, 2479 (1999).

\bibitem{VBS00}
M.~Vojta, C.~Buragohain and S.~Sachdev,
Phys. Rev. B {\bf 61}, 15152 (2000).

\bibitem{si01}
Q. Si, S. Rabello, K. Ingersent, and J. L. Smith,
Nature (London) \textbf{413}, 804 (2001).

\bibitem{si03}
Q. Si, S. Rabello, K. Ingersent, and J.~L. Smith,
Phys. Rev. B \textbf{68}, 115103 (2003).

\bibitem{SY93}
S. Sachdev and J. Ye,
Phys. Rev. Lett. \textbf{70}, 3339 (1993).

\bibitem{SYKreview}
D. Chowdhury, A Georges, O. Parcollet, and S. Sachdev,
Rev. Mod. Phys. \textbf{94}, 035004 (2022).

\bibitem{mv06}
M. Vojta,
Philos. Mag. \textbf{86}, 1807 (2006).

\bibitem{guo12}
C. Guo, A. Weichselbaum, J. von Delft, and M. Voj\-ta,
Phys. Rev. Lett. {\bf 108}, 160401 (2012).

\bibitem{bruo14}
B. Bruognolo, A. Weichselbaum, C. Guo, J. von Delft, I. Schneider, and M. Voj\-ta,
Phys. Rev. B {\bf 90}, 245130 (2014).

\bibitem{weber21}
M. Weber,
Phys. Rev. B \textbf{105}, 165129 (2022).

\bibitem{bfk}
J.~L.~Smith and Q.~Si, cond-mat/9705140; Europhys. Lett. {\bf 45}, 228 (1999).

\bibitem{sengupta}
A.~M.~Sengupta, Phys. Rev. B {\bf 61}, 4041 (2000).

\bibitem{kehrein96}
S.~K. Kehrein and A.~Mielke,
Phys. Lett. A {\bf 219}, 313 (1996).

\bibitem{bulla03}
R.~Bulla, N.~H.~Tong, and M.~Vojta,
Phys. Rev. Lett. {\bf 91}, 170601 (2003).

\bibitem{VTB05}
M.~Vojta, N.~H.~Tong, and R.~Bulla,
Phys. Rev. Lett. {\bf 94}, 070604 (2005);
Phys. Rev. Lett. {\bf 102}, 249904(E) (2009).

\bibitem{rieger09}
A. Winter, H. Rieger, M. Vojta, and R. Bulla,
Phys. Rev. Lett. {\bf 102}, 030601 (2009).

\bibitem{fisher_critical_1972}
M.~E. Fisher, S.~K. Ma, and B.~G. Nickel,
Phys. Rev. Lett. {\bf 29}, 917 (1972).

\bibitem{bulla05}
R.~Bulla, H.~J.~Lee, N.~H.~Tong, and M.~Vojta,
Phys. Rev. B {\bf 71}, 045122 (2005).

\bibitem{otsuki13}
J. Otsuki,
Phys. Rev. B \textbf{87}, 125102 (2013).

\bibitem{cai19}
A. Cai and Q. Si,
Phys. Rev. B \textbf{100}, 014439 (2019).

\bibitem{rg_bfk}
L. Zhu and Q. Si, Phys. Rev. B {\bf 66}, 024426 (2002);
G. Zarand and E. Demler, Phys. Rev. B {\bf 66}, 024427 (2002).

\bibitem{ren_note}
$\alpha$ and $\bar\alpha$ in Eqs.~\eqref{beta_wk}, \eqref{alstar}, \eqref{beta_str}, and \eqref{eq:alqc} represent renormalized couplings in the RG framework (as opposed to bare couplings; for notational simplicity, we use the same symbol for both). Their definition depends on the RG scheme; our choice differs by a factor of 2 from that of Refs.~\cite{rg_bfk} to match $\alpha_\mathrm{CR}$ to the microscopic calculation.

\bibitem{weber17}
M. Weber, F. F. Assaad, and M. Hohenadler,
Phys. Rev. Lett. \textbf{119}, 097401 (2017).


\bibitem{SupplInfo}
See Supplemental Material for
additional results for the local-moment formation,
details on our finite-size scaling analysis, and
a numerical confirmation of hyperscaling, which includes
Refs.~\cite{FSS14,toldin15,bruoThesis}.


\bibitem{FSS14}
M. Campostrini, A. Pelissetto, and E. Vicari,
Phys. Rev. B {\bf 89}, 094516 (2014).

\bibitem{toldin15}
F. Parisen Toldin, M. Hohenadler, F. F. Assaad, and I. F. Herbut,
Phys. Rev. B {\bf 91}, 165108 (2015).

\bibitem{bruoThesis}
B. Bruognolo,
Master's Thesis, LMU Munich, 2013.




\bibitem{Sandvik10}
A. W. Sandvik,
AIP Conf. Proc. \textbf{1297}, 135 (2010).

\bibitem{sandvik07}
K. H. Hoglund and A. W. Sandvik,
Phys. Rev. Lett. \textbf{99}, 027205 (2007).

\bibitem{cuoma22}
G. Cuomo, Z. Komargodski, M. Mezei, and A. Raviv-Moshe,
J. High Energy Phys. 06 (2022) 112.

\bibitem{beccaria22}
M. Beccaria, S. Giombi, and A. Tseytlin,
J. Phys. {\bf 55}, 255401 (2022).

\bibitem{Nahum22}
A. Nahum,
Phys. Rev. B {\bf 106}, L081109 (2022).




\end{thebibliography}
\end{document}